\documentclass[twoside,twocolumn,9pt]{article}
\usepackage{extsizes}
\usepackage[super,sort&compress,comma]{natbib} 
\usepackage[version=3]{mhchem}
\usepackage[left=1.5cm, right=1.5cm, top=1.785cm, bottom=2.0cm]{geometry}
\usepackage{balance}
\usepackage{mathptmx}
\usepackage{sectsty}
\usepackage{graphicx} 
\usepackage{lastpage}
\usepackage[format=plain,justification=justified,singlelinecheck=false,font={stretch=1.125,small,sf},labelfont=bf,labelsep=space]{caption}
\usepackage{float}
\usepackage{fancyhdr}
\usepackage{fnpos}
\usepackage[english]{babel}
\addto{\captionsenglish}{%
  
}
\usepackage{array}
\usepackage{droidsans}
\usepackage{charter}
\usepackage[T1]{fontenc}
\usepackage[usenames,dvipsnames]{xcolor}
\usepackage{setspace}
\usepackage[compact]{titlesec}
\usepackage{hyperref}
\usepackage{mhchem}

\usepackage{epstopdf}

\definecolor{cream}{RGB}{222,217,201}

\begin{document}

\pagestyle{fancy}
\thispagestyle{plain}
\fancypagestyle{plain}{
\renewcommand{\headrulewidth}{0pt}
}

\newcommand{\Rdust}{$R_\mathrm{dust}$}
\newcommand{\Tcomp}{$T_\mathrm{comp}$}
\newcommand{\Rcomp}{$R_\mathrm{comp}$}
\newcommand{\msunyr}{M$_\odot$ yr$^{-1}$}
\newcommand{\msun}{M$_\odot$}
\newcommand{\mdot}{$\dot{M}$}
\newcommand{\kms}{km s$^{-1}$}
\newcommand{\fvol}{$f_\mathrm{vol}$}
\newcommand{\fic}{$f_\mathrm{ic}$}
\newcommand{\amin}{$a_\mathrm{min}$}
\newcommand{\amax}{$a_\mathrm{max}$}

\makeFNbottom
\makeatletter
\renewcommand\LARGE{\@setfontsize\LARGE{15pt}{17}}
\renewcommand\Large{\@setfontsize\Large{12pt}{14}}
\renewcommand\large{\@setfontsize\large{10pt}{12}}
\renewcommand\footnotesize{\@setfontsize\footnotesize{7pt}{10}}
\makeatother

\renewcommand{\thefootnote}{\fnsymbol{footnote}}
\renewcommand\footnoterule{\vspace*{1pt}%
\color{cream}\hrule width 3.5in height 0.4pt \color{black}\vspace*{5pt}} 
\setcounter{secnumdepth}{5}

\makeatletter 
\renewcommand\@biblabel[1]{#1}            
\renewcommand\@makefntext[1]%
{\noindent\makebox[0pt][r]{\@thefnmark\,}#1}
\makeatother 
\renewcommand{\figurename}{\small{Fig.}~}
\sectionfont{\sffamily\Large}
\subsectionfont{\normalsize}
\subsubsectionfont{\bf}
\setstretch{1.125} 
\setlength{\skip\footins}{0.8cm}
\setlength{\footnotesep}{0.25cm}
\setlength{\jot}{10pt}
\titlespacing*{\section}{0pt}{4pt}{4pt}
\titlespacing*{\subsection}{0pt}{15pt}{1pt}

\fancyfoot{}
\fancyfoot[LO,RE]{\vspace{-7.1pt}\includegraphics[height=9pt]{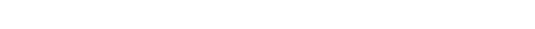}}
\fancyfoot[CO]{\vspace{-7.1pt}\hspace{13.2cm}\includegraphics{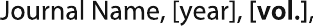}}
\fancyfoot[CE]{\vspace{-7.2pt}\hspace{-14.2cm}\includegraphics{head_foot/RF}}
\fancyfoot[RO]{\footnotesize{\sffamily{1--\pageref{LastPage} ~\textbar  \hspace{2pt}\thepage}}}
\fancyfoot[LE]{\footnotesize{\sffamily{\thepage~\textbar\hspace{3.45cm} 1--\pageref{LastPage}}}}
\fancyhead{}
\renewcommand{\headrulewidth}{0pt} 
\renewcommand{\footrulewidth}{0pt}
\setlength{\arrayrulewidth}{1pt}
\setlength{\columnsep}{6.5mm}
\setlength\bibsep{1pt}

\makeatletter 
\newlength{\figrulesep} 
\setlength{\figrulesep}{0.5\textfloatsep} 

\newcommand{\topfigrule}{\vspace*{-1pt}%
\noindent{\color{cream}\rule[-\figrulesep]{\columnwidth}{1.5pt}} }

\newcommand{\botfigrule}{\vspace*{-2pt}%
\noindent{\color{cream}\rule[\figrulesep]{\columnwidth}{1.5pt}} }

\newcommand{\dblfigrule}{\vspace*{-1pt}%
\noindent{\color{cream}\rule[-\figrulesep]{\textwidth}{1.5pt}} }

\ExplSyntaxOn
\keys_define:nn { mhchem }
 {
  arrow-min-length .code:n =
   \cs_set:Npn \__mhchem_arrow_options_minLength:n { {#1} } 
 }
\ExplSyntaxOff

\makeatother

\twocolumn[
  \begin{@twocolumnfalse}
{\includegraphics[height=30pt]{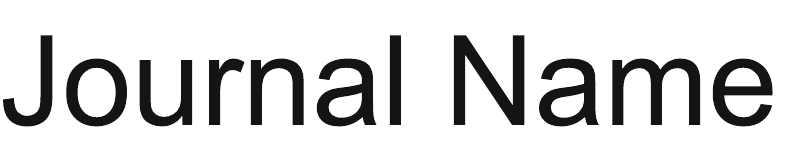}\hfill\raisebox{0pt}[0pt][0pt]{\includegraphics[height=55pt]{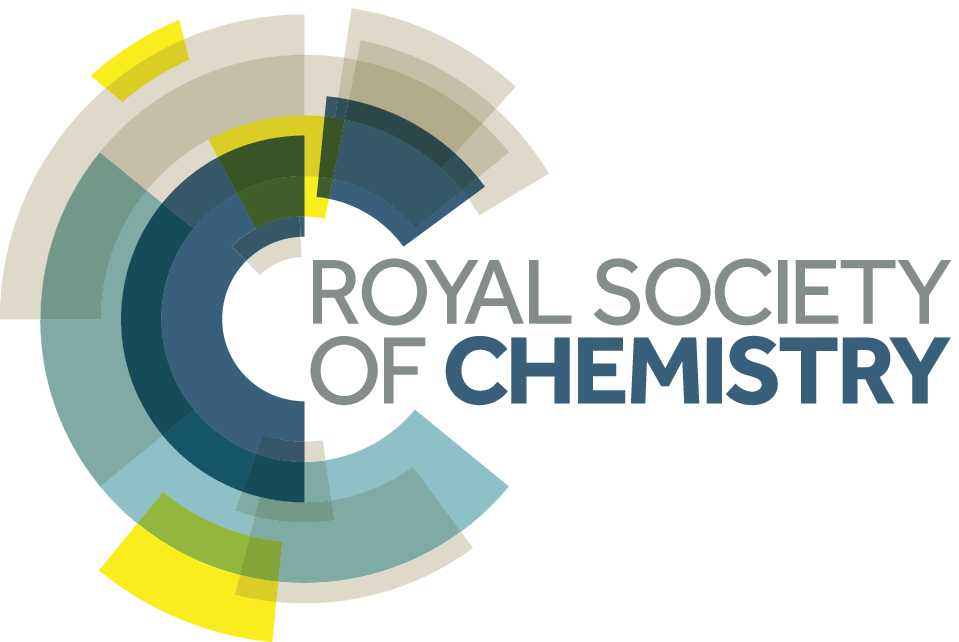}}\\[1ex]
\includegraphics[width=18.5cm]{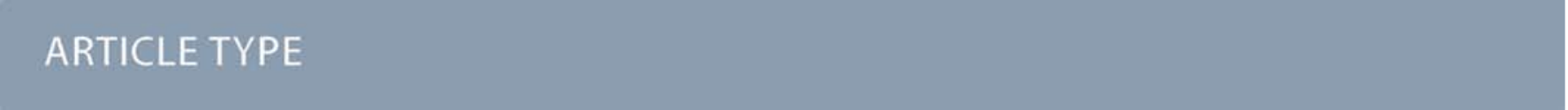}}\par
\vspace{1em}
\sffamily
\begin{tabular}{m{4.5cm} p{13.5cm} }

\includegraphics{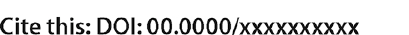} & \noindent\LARGE{\textbf{Disentangling physics and chemistry in AGB outflows: revealing degeneracies when adding complexity$^\dag$}} \\
\vspace{0.3cm} & \vspace{0.3cm} \\

 & \noindent\large{Marie Van de Sande,$^{\ast}$\textit{$^{a}$} Catherine Walsh,\textit{$^{a}$} and Tom J. Millar\textit{$^{b}$}} \\

\includegraphics{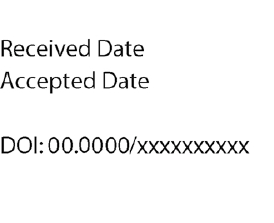} & \noindent\normalsize{Observations of the outflows of asymptotic giant branch (AGB) stars continue to reveal their chemical and dynamical complexity. 
Spherical asymmetries, such as spirals and disks, are prevalent and thought to be caused by binary interaction with a (sub)stellar companion.
Furthermore, high density outflows show evidence of dust-gas interactions.
The classical chemical model of these outflows --- a gas-phase only, spherically symmetric chemical kinetics model --- is hence not appropriate for a majority of observed outflows.
We have included several physical and chemical advancements step-by-step: a porous density distribution, dust-gas chemistry, and internal UV photons originating from a close-by stellar companion.
Now, we combine these layers of complexity into the most chemically and physically advanced chemical kinetics model of AGB outflows to date.
By varying over all model parameters, we obtain a holistic view of the outflow's composition and how it (inter)depends on the different complexities.
A stellar companion has the largest influence, especially when combined with a porous outflow.
We compile sets of gas-phase molecules that trace the importance of dust-gas chemistry and allow us to infer the presence of a companion and porosity of the outflow.
This shows that our new chemical model can be used to infer physical and chemical properties of specific outflows, as long as a suitable range of molecules is observed. 
} \\



\end{tabular}

 \end{@twocolumnfalse} \vspace{0.6cm}

  ]

\renewcommand*\rmdefault{bch}\normalfont\upshape
\rmfamily
\section*{}
\vspace{-1cm}


\footnotetext{\textit{$^{a}$~School of Physics and Astronomy, University of Leeds, Leeds LS2 9JT, UK; E-mail: m.vandesande@leeds.ac.uk}}
\footnotetext{\textit{$^{b}$~Astrophysics Research Centre, School of Mathematics and Physics, Queen’s University Belfast, University Road, Belfast BT7 1NN, UK. }}

\footnotetext{\dag~Electronic Supplementary Information (ESI) available: fractional abundance profiles of all parents and their daughters along with all species that show a significant change in abundance when including a stellar companion, together with figures showing the number of monolayers of dust and ice for all assumed GSDs and the ice and refractory coverage at the end of the outflow for the GSDs different to the MRN distribution.
See DOI: 00.0000/00000000.}




\mhchemoptions{arrow-min-length=1em}

\section{Introduction}			\label{sect:intro}

During the asymptotic giant branch (AGB) phase, stars with an initial mass up to $8$ \msun\ lose their outer layers by means of a stellar outflow or wind.
This process of mass loss is efficient, with mass-loss rates between $10^{-8}$ and $10^{-4}$ \msunyr, and deterimines the star's remaining lifetime rather than exhaustion of nuclear fuel in its core.
The outflow is thought to be launched by a two-step mechanism: stellar pulsations facilitate the formation of dust grains, subsequently launching a dust-driven wind \cite{Habing2003,Hofner2018}. 
Thanks to their outflows, AGB stars are important contributors to the chemical enrichment of the interstellar medium (ISM), contributing about $80\%$ of the stellar gas injection rate \cite{Tielens2005} and about $70\%$ of the total stellar dust production rate \cite{Zhukovska2013}.

The chemical content of the outflow is determined by the elemental carbon-to-oxygen ratio (C/O) of the AGB star, linked to its initial mass and evolutionary stage. 
Stars with C/O $< 1$ give rise to an oxygen-rich (O-rich) outflow, while stars with C/O $> 1$ have a carbon-rich (C-rich) outflow.
The outflows of AGB stars are rich astrochemical laboratories: close to 100 different molecules and some 15 types of newly formed dust species have been detected so far.
Moreover, they are host to different types of chemistry thanks to the large gradients in density and temperature present throughout the outflow.
Three main regions can be distinguished: non-equilibrium chemistry caused by shocks due to stellar pulsations close to the star, followed by dust condensation and dust-gas interaction, to photochemistry initiated by interstellar UV photons in the tenuous outer region \cite{Decin2021}.

Chemical models of AGB outflows are divided into two main groups, dealing with the non-equilibrium chemistry close to the star  \cite{Koehler1997,Cherchneff2006,Goumans2012,Gobrecht2016} or with the photochemistry of the outer wind \cite{Huggins1982,Nejad1984,Millar1994,Li2016}.
Both types of models include gas-phase chemistry only and assume a spherically symmetric outflow.
Observations continue to reveal discrepancies with model predictions, putting an increasingly larger strain on these two assumptions.
Spherically asymmetric outflows are common: both small-scale inhomogeneities or clumps \cite{Leao2006,Khouri2016,Agundez2017} and large-scale structures such as spirals \cite{Mauron2006,Maercker2016} and disks \cite{Kervella2014,Homan2020} are widely observed. 
These large-scale structures are thought to be caused by binary interaction with a (sub)stellar companion.
Most AGB stars with mass-loss rates larger than $\sim 10^{-7}$ \msunyr\ will have at least one planetary and/or stellar companion\cite{Decin2020}. The potential presence of another star complicates the chemical picture even further as it produces an internal UV radiation field.

Moreover, in both O-rich and C-rich outflows, the abundance of refractory species (e.g., SiO, SiS, and CS) is observed to decrease with increasing outflow density, hinting towards depletion onto dust grains \cite{GonzalezDelgado2003,Massalkhi2019}.
The abundance of SiO and SiS is observed to decrease in the intermediate wind of high density O-rich outflows, before the onset of dust formation \cite{Bujarrabal1989,Decin2010a}, and \ce{H2O} ice has been detected around OH/IR stars \cite{Sylvester1999}.

\begin{table}[t]
\small
  \caption{\ Physical parameters of the grid of chemical models}
  \label{table:model}
  \begin{tabular*}{0.48\textwidth}{@{\extracolsep{\fill}}ll}
    \hline
    \noalign{\smallskip}
	\multicolumn{2}{c}{Fixed parameters} \\
    \hline
    \noalign{\smallskip}
    \multicolumn{2}{c}{Physical parameters} \\
    \noalign{\smallskip}
    Stellar radius, $R_*$             & 5 $\times 10^{13}$ cm \\
    Stellar temperature, $T_*$        & $2330$ K \\
    Exponent $T(r)$, $\epsilon$                    & $0.7$ \\

    \noalign{\smallskip}
    \multicolumn{2}{c}{Dust parameters} \\
    \noalign{\smallskip}
    O-rich dust composition	& 50/50 melilite and silica with Fe \\
    C-rich dust composition	& Amorphous carbon \\  
	Dust-to-gas mass ratio			& $2 \times 10^{-3}$ \\
	Surface density of binding sites	& $10^{15}$ cm$^{-2}$ \\
	Silicate dust bulk density\cite{Draine2003}	& 3.5 g cm$^{-3}$  \\
	Carbonaceous dust bulk density\cite{Draine2003} 	& 2.24 g cm$^{-3}$  \\
    \noalign{\smallskip}

    \hline
    \noalign{\smallskip}
	\multicolumn{2}{c}{Variable parameters} \\
    \hline
    \noalign{\smallskip}
    \multicolumn{2}{c}{Physical parameters} \\
    \noalign{\smallskip}    
    Initial radius		& $1.025 \times R_\mathrm{dust}$ \\
	\noalign{\smallskip}								
    Grid of outflow densities & \\
	\noalign{\smallskip}								
        \quad Mass-loss rate $\dot{M}$;        					&    $10^{-5}$ \msunyr; $15$ \kms; $5$ \kms \\
    	\quad expansion velocity, $v_\infty$;				&    $10^{-6}$ \msunyr; $10$ \kms; $10$ \kms  \\
    	\quad and drift velocity, $v_\mathrm{drift}$		&    $10^{-7}$ \msunyr; $5$ \kms; $15$ \kms \\
	\noalign{\smallskip}								
	Grid of density structures		& \\
	\noalign{\smallskip}								
	 \quad Smooth outflow &  \\
	 \quad Porous outflows, with & \\
	 \quad \quad interclump density contrast	&  \fic\ $=\ 0.0,\ 0.1,\ 0.3,\ 0.5$ \\
	 \quad \quad clump volume filling factor	&  \fvol\ $=\ 0.1,\ 0.3,\ 0.5$ \\
	 \quad \quad size of the clumps at $R_*$	&   $l_*\ =\ 5\times 10^{12},\ 10^{13}$ cm \\
    \noalign{\smallskip}
    \multicolumn{2}{c}{Dust parameters} \\
    \noalign{\smallskip}
	Grain size distributions & \\
	\noalign{\smallskip}								
    \quad Minimum grain size, \amin;	& 				$10^{-8}$ cm; $10^{-6}$ cm; $-3.5$	\\
    \quad maximum grain size, \amax; & 		$5 \times 10^{-7}$ cm; $2.5 \times 10^{-5}$ cm; $-3.5$ 	\\ 
	\quad and power-law exponent, $\eta$ & 	$10^{-6}$ cm; $10^{-4}$ cm; $-4.5$	\\
    \noalign{\smallskip}
    \multicolumn{2}{c}{Companion parameters} \\
    \noalign{\smallskip}
    Onset of dust extinction, $R_\mathrm{dust}$		& $2,\ 5$ $R_*$ \\
    Companion temp., \Tcomp; 					& $4000$ K; $1.53\times 10^{10}$ cm \\
    \quad and radius, \Rcomp					& $6000$ K; $8.14\times 10^{10}$ cm  \\
												& $10\ 000$ K; $6.96 \times 10^{8}$ cm \\

	\hline
	    \noalign{\smallskip}

  \end{tabular*}
\end{table}

To better explain observations, we have increased the physical and chemical complexity of earlier gas-phase only models through the inclusion of: a clumpy density distribution \citep{VandeSande2018}, dust-gas chemistry \citep{VandeSande2019b,VandeSande2020,VandeSande2021}, and internal UV photons originating from a close-by stellar companion \citep{VandeSande2019a,VandeSande2022}.
This was done step-by-step, adding separate layers of complexity to investigate their effects on the composition of the outflow in isolation.
While clumpy outflows were considered when including internal UV photons, the interdependencies of the other complexities have not yet been explored.
To this end, we have combined all previous developments in the most physically and chemically complex chemical model of AGB outflows to date.
This increases the model's resolution by extending the chemical validity of the model from the outer wind into the dust interaction zone and increasing its applicability to spherically asymmetric outflows.

Exploring the interplay of the model developments enables us to determine how the chemistry of the outflow depends on its structure and the presence of a (sub)stellar companion.
By maintaining a holistic view of the chemical composition throughout the outflow and the different chemical and physical processes included, we are able to determine a suite of molecules that indicate the presence of (a combination of) these complexities.
This makes the chemical model a powerful tool to interpret observations and reveal the hidden cause of dynamical structures.

We describe the chemical model in Sect. \ref{sect:model}, along with the different complexities and their limitations.
We present the results for C-rich and O-rich outflows in Sect. \ref{sect:results}.
Discussion and conclusions follow in Sects \ref{sect:disc} and \ref{sect:conclusions}.



\section{AGB outflow chemical model}		\label{sect:model}

The one-dimensional chemical kinetics model is based on the publicly available UDfA circumstellar envelope model \citep{McElroy2013}\footnote{\url{http://udfa.ajmarkwick.net/index.php?mode=downloads}}. 
It describes a spherically symmetric outflow with constant mass-loss rate, $\dot{M}$, and expansion velocity, $v_\infty$.
The temperature profile of the gas is described by a power law\citep{VandeSande2018}
\begin{equation}
 T(r) =  T_* \left(\frac{r}{R_*}\right)^{-\epsilon},
\end{equation}
with $T_*$ and $R_*$ the stellar temperature and radius, respectively, and $r$ the distance from the centre of the star.
CO self-shielding is taken into account using a single-band approximation \citep{Morris1983}; \ce{H2} is assumed to be fully self-shielded.

Table \ref{table:model} lists the different fixed and variable model parameters along with their values adopted in this work.
For each configuration, we calculate an O-rich and a C-rich outflow.
The parent species for the O-rich and C-rich outflows are listed in Table \ref{table:model-parents}.
These species are assumed to be present at the start of the model; they are taken from Ag\'undez et al.\,(2020) \cite{Agundez2020}, who compiled (ranges of) observed abundances in the inner region. 
We consider three outflow densities, determined by the mass-loss rate, $\dot{M}$ and the expansion velocity, $v_\infty$. 
The value of $v_\infty$ and the drift velocity $v_\mathrm{drift}$ for each mass-loss rate is estimated from observations \cite{Schoier2001,Olofsson2002,GonzalezDelgado2003,Ramstedt2008,Danilovich2015}.

\begin{figure*}[ht]
 \centering
 \includegraphics[width=17cm]{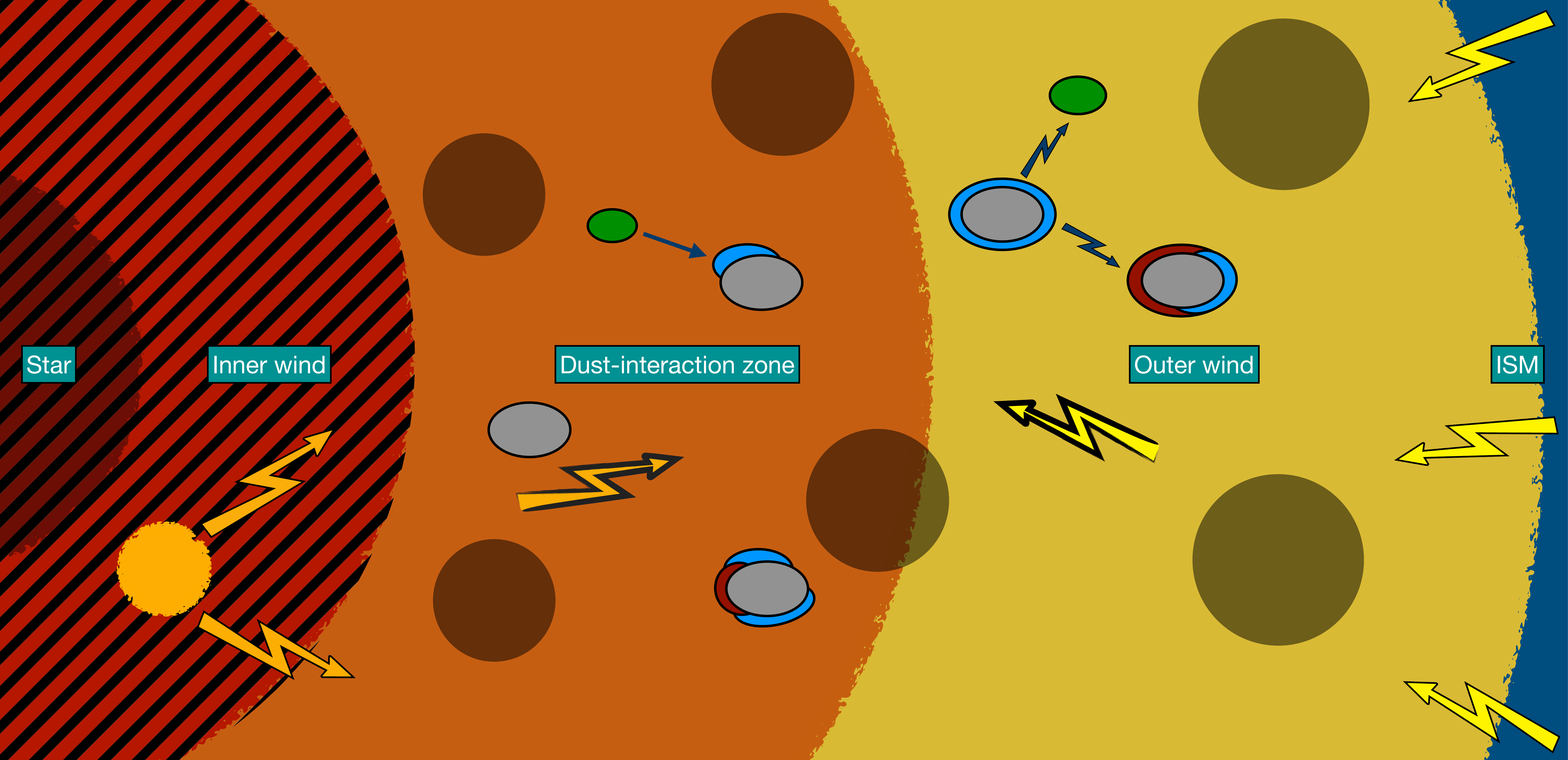}
 \caption{Overview of an AGB star and its outflow with the different complexities included in the chemical model.
 The model does not describe the chemistry the cross-hatched inner wind region. 
 Interstellar UV radiation always irradiates the outer regions of the outflow (yellow lightning bolts).
 A \emph{companion star} can be present within the inner wind region, resulting in an inner wind UV radiation field (orange lightning bolts).
 Including \emph{porosity} distributes the outflow's material between overdense clumps (opaque black circles) and a rarified interclump medium.
 This affects extinction throughout the outflow, with interstellar photons reaching closer to the star and potential companion photons reaching further out in the outflow (lightning bolts with black outlines).
 \emph{Dust-gas chemistry} leads to the accretion of gas-phase species (green ovals) onto dust grains (grey ovals), forming an ice mantle on top of the dust (blue).
 Ice species, which may be formed via grain-surface chemistry, are released into the gas phase by photodesorption.
 If the ice is complex, it can be photoprocessed into refractory organic material (red), both by interstellar and companion UV photons.
}
 \label{fig:network}
\end{figure*}

We outline important features and parameters of the physical and chemical developments along with our previous main findings in Sects \ref{model:phys} and \ref{model:chem}, respectively.
For additional details, we refer to the original works.
In Sect. \ref{subsect:model:limits}, we elaborate on the assumptions made when implementing the developments and the limitations of the model.
The different complexities included in the chemical model are illustrated in Fig. \ref{fig:network}.

\begin{table}[t]
	\caption{Parent species and their abundances relative to \ce{H2} for the C-rich and O-rich outflows.
	Abundances are derived from observations, as compiled by \citet{Agundez2020}. 
	} 
    \centering
    \begin{tabular}{l r c  l r }
    \hline  
    \multicolumn{2}{c}{Carbon-rich} && \multicolumn{2}{c}{Oxygen-rich}  \\  
    \cline{1-2} \cline{4-5} 
    \noalign{\smallskip}
    Species & Abun. & & Species & Abun. \\
    \cline{1-2} \cline{4-5} 
    \noalign{\smallskip}
    He		&  0.17				& & He		& 0.17  \\
    CO		& $8.00\times10^{-4}$	& & CO		& $3.00 \times 10^{-4}$  \\
    N$_2$		& $4.00 \times 10^{-5}$	& & H$_2$O	& $2.15 \times 10^{-4}$  \\
    CH$_4$	& $3.50 \times 10^{-6}$	& & N$_2$ 	& $4.00 \times 10^{-5}$  \\ 
    H$_2$O	& $2.55 \times 10^{-6}$	& & SiO 	& $2.71 \times 10^{-5}$  \\ 
    SiC$_2$	& $1.87 \times 10^{-5}$	& & H$_2$S 	& $1.75 \times 10^{-5}$  \\
    CS		& $1.06 \times 10^{-5}$	& & SO$_2$ 	& $3.72 \times 10^{-6}$  \\
    C$_2$H$_2$& $4.38 \times 10^{-5}$	& & SO 		& $3.06 \times 10^{-6}$  \\
    HCN		& $4.09 \times 10^{-5}$	& & SiS 		& $9.53 \times 10^{-7}$  \\
    SiS   		& $5.98 \times 10^{-6}$	& & NH$_3$ 	& $6.25 \times 10^{-7}$  \\ 
    SiO 		& $5.02 \times 10^{-6}$	& & CO$_2$ 	& $3.00 \times 10^{-7}$  \\   
    HCl		& $3.25 \times 10^{-7}$	& & HCN 	& $2.59 \times 10^{-7}$  \\  
    C$_2$H$_4$& $6.85 \times 10^{-8}$	& & PO 		& $7.75 \times 10^{-8}$  \\ 
    NH$_3$	& $6.00 \times 10^{-8}$	& & CS 		& $5.57 \times 10^{-8}$  \\
    HCP		& $2.50 \times 10^{-8}$	& & PN 		& $1.50 \times 10^{-8}$  \\
    HF    		& $1.70 \times 10^{-8}$	& &  HCl		& $1.00 \times 10^{-8}$  \\  
    H$_2$S	& $4.00 \times 10^{-9}$	& & 	HF	& $1.00 \times 10^{-8}$  \\    
    \hline 
    \end{tabular}%
    \label{table:model-parents}    
\end{table}

\subsection{Physical complexity}			\label{model:phys}

Clumps are taken into account by using a porosity formalism, which divides the outflow into a stochastic two-component medium of overdense clumps within a rarified interclump component \citep{Owocki2004,Owocki2006,Sundqvist2012,Sundqvist2014}.
The formalism allows for a one-dimensional approximation of a clumpy outflow by including the effects of the local overdensity within the clumps along with their effect on the outflow's optical depth.

The specific clumpiness of the outflow is described by three parameters: (i) the clump volume filling factor, \fvol, setting the fraction of the total volume of the outflow occupied by clumps, (ii) the interclump density contrast, \fic, setting the fraction of the outflow density within the interclump component, and (iii) the size of the clumps at the stellar surface, $l_*$.
Clump size and volume filling factor can be combined into the porosity length $h = l_*/f_\mathrm{vol}$, which represents the local mean free path between clumps.
Highly porous outflows have a large porosity length, due to a large $l_*$ and/or small \fvol, combined with a low interclump density contrast.

In our previous work \citep{VandeSande2018,VandeSande2018Err}, which included gas-phase only chemistry and no internal UV photons, we found that the effects of the local overdensity of clumps is small, while those of the change in optical depth can be considerable. 
With increasing porosity, parent species are photodissociated closer to the star, though still in the outer region. 
This leads to larger abundances of daughter species as well as a shift in their peak abundances closer to the star.
However, these changes are of the order of one magnitude or less.

\subsection{Chemical complexity}			\label{model:chem}

The chemical reaction network used is a combination of the dust-gas chemical network used in Van de Sande et al.\,(2021)\citep{VandeSande2021} and the inner UV photon reactions of Van de Sande \& Millar\,(2022)\cite{VandeSande2022}. 
We elaborate on these networks and their implementation below.

\subsubsection{Dust-gas chemistry}

The comprehensive dust-gas chemical reaction network is that of Walsh et al. (2014)\citep{Walsh2014}, which includes the original UDfA {\sc{Rate12}} gas-phase only chemical network\cite{McElroy2013} extended with grain-surface reactions and related reactions such as needed gas-phase reactions from the 2008 version of the Ohio State University network\citep{Garrod2008}. 
Additionally, we included the photoprocessing of complex ices into refractory organic material as described in Van de Sande et al. (2021)\citep{VandeSande2021}.
By including these dust-gas reactions, the chemical validity of the model is extended from the outer wind into the intermediate wind.

The network allows for gas-phase species to accrete onto dust grains, forming a physisorbed ice mantle which can be returned to the gas-phase via thermal desorption, photodesorption, and mechanical sputtering. 
Reactions in the ice mantle can occur via the diffusive Langmuir-Hinshelwood and stick-and-hit Eley-Rideal mechanisms, which both can give rise to reactive desorption.
Icy species can be photodissociated, with the products either remaining on the dust or being returned to the gas-phase. 
The photoprocessing of ices is limited to complex species only, which we define as species containing at least three C atoms or two C atoms and another heavy atom.

As dust nucleation is not included in the network, dust grains are assumed to be present at the start of the model with a constant density.
The dust has a drift velocity, $v_\mathrm{drift}$ relative to the gas, defined as $v_\mathrm{drift} = v_\mathrm{dust} - v_\infty$.
The grain size distribution (GSD) is fixed and parametrised following the MRN grain-size distribution\cite{Mathis1977,VandeSande2020}, where the dust grain number density per unit volume is given by
\begin{equation}			\label{eq:MRN}
n_\mathrm{dust} = \frac{C}{r^2}\ \int_{a_\mathrm{min}}^{a_\mathrm{max}} a^\eta\ {d}a\ \mathrm{cm^{-3}},
\end{equation}
where \amin\ and \amax\ are the minimum and maximum grain size and $\eta$ is the slope of the power law. 
We vary the GSD over the canonical MRN distribution, with \amin\ $= 5 \times 10^{-7}$ cm, \amax\ $ = 2.5 \times 10^{-5}$, and $\eta = -3.5$ \cite{Mathis1977}, and GSDs with smaller and larger grains than the MRN distribution (Table \ref{table:model}).

The dust's temperature is approximated by a power law
\begin{equation} 		\label{eq:Tdust}
T_\mathrm{dust} = T_\mathrm{dust,*} \left(	\frac{2r}{R_*} \right)^{{2}/{(4+s)}},
\end{equation}
where $s$ and $T_\mathrm{dust,*}$ are free parameters. 
These parameters are constrained by fitting Eq. (\ref{eq:Tdust}) to the results of a continuum radiative transfer model for different outflow densities and dust compositions. 
We assume the same dust composition as in Van de Sande et al. (2021)\cite{VandeSande2021}: a 50/50 mixture of melilite and silicate with Fe for O-rich outflows and amorphous carbon for C-rich outflows.

In our previous work\cite{VandeSande2019b,VandeSande2020,VandeSande2021}, we found that dust-gas chemistry can have a large impact on higher density outflows ($\dot{M} \geq 10^{-6}$ \msunyr).
Parent species with higher binding energies are depleted onto dust grain, while daughter species can be formed in the ice mantle and released by photodesorption in the outer regions.\cite{VandeSande2019b}
The depletion level is influenced by the GSD, where GSDs with larger average surface areas (caused by a smaller \amin and/or steeper slopes) lead to larger depletion of parent species.\cite{VandeSande2020}
The formation of an ice mantle mainly depends on outflow density and the chemistry of the outflow, with up to 100 monolayers of \ce{H2O}-rich ice building up in very high density O-rich outflows and sub-monolayer coverage only ($\sim 0.1$ monolayers) in C-rich outflows.\cite{VandeSande2019b}
Refractory organic material is more readily formed in C-rich outflows thanks to a larger abundance of C-rich feedstock ices, forming up to 0.2 monolayers, compared to negligible coverage ($\sim10^{-5}$ only) for O-rich outflows.

\begin{figure*}[ht]
 \centering
 \includegraphics[width=18cm]{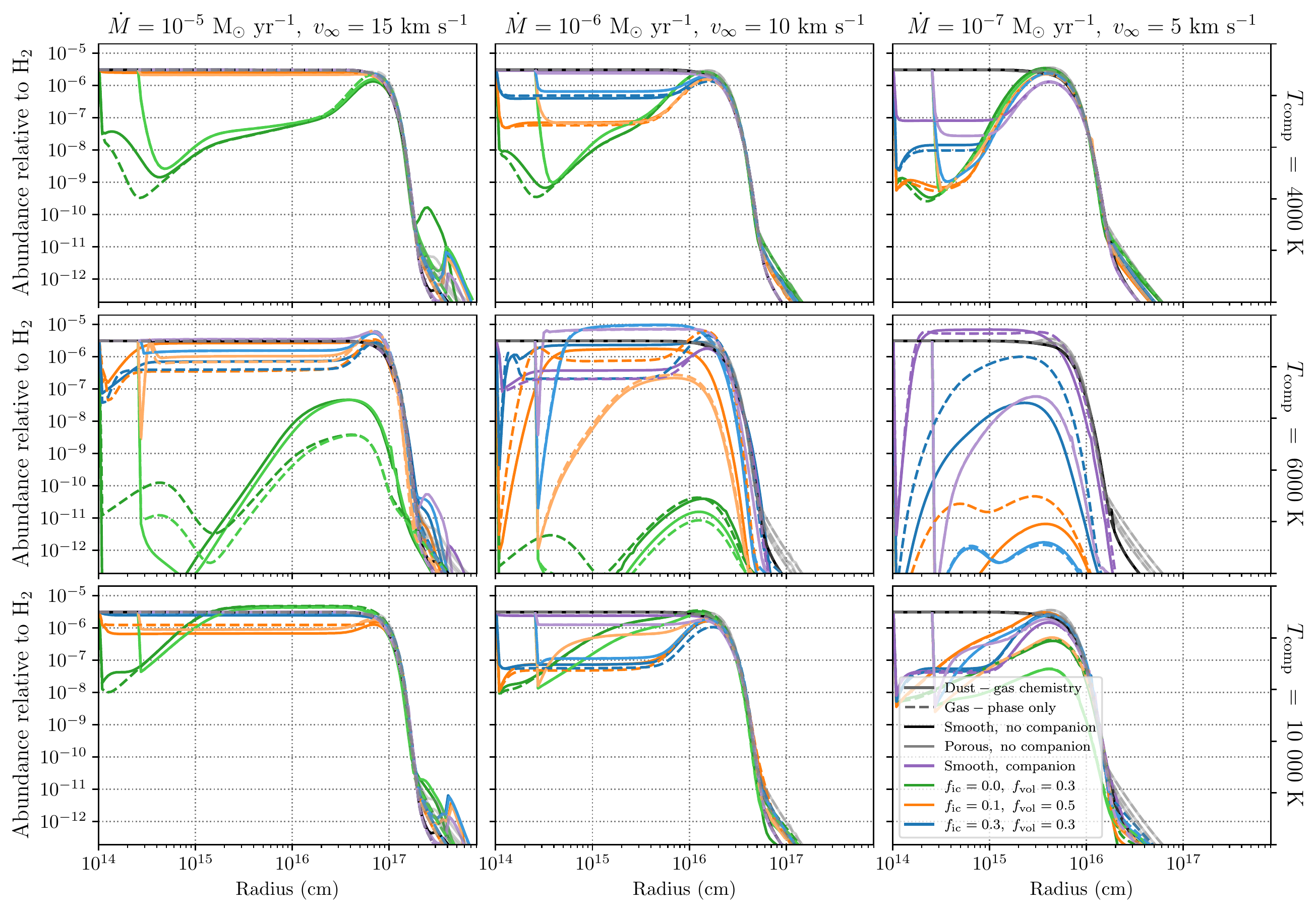}
 \caption{Fractional abundance of the parent \ce{SO} relative to \ce{H2} for a selection of O-rich outflows with different outflow densities (columns) and companions (rows). 
 Different line styles show different chemistries included. Dashed lines: gas-phase chemistry only, solid lines: including dust-gas chemistry.
 Different colours show different density structures and the presence of a companion.
 Black: smooth outflow without a companion, gray: porous outflow without a companion.
 When including a companion, different shades correspond to the value of \Rdust. 
 Darker shade: \Rdust $= 2\ R_*$, lighter shade: \Rdust $= 5\ R_*$.
 Purple: smooth outflow with a companion. 
 Green: porous outflow with a companion, \fic $= 0.0$, \fvol $= 0.3$, $l_* = 10^{13}$ cm.
 Orange: porous outflow with a companion, \fic $= 0.1$, \fvol $= 0.5$, $l_* = 10^{13}$ cm.
 Blue: porous outflow with a companion, \fic $= 0.3$, \fvol $= 0.3$, $l_* = 5\times10^{12}$ cm.
}
 \label{fig:Orich-SO}
\end{figure*}

\subsubsection{Internal UV photons}			\label{model:chem:companion}

The binary fraction of AGB stars strongly depends on the their initial mass.
For AGB stars with an initial mass smaller than $1.5$ M$_\odot$, $\sim 33\%$ have a stellar companion and $\sim 10\%$ a planetary or brown dwarf companion capable of shaping the outflow.
However, these stars only have a mass-loss rate larger $10^{-7}$ \msunyr, the smallest value in our model grid , for a short period.
Therefore, observed AGB outflows with $\dot{M} \geq 10^{-7}$ \msunyr\ will generally have an initial mass larger than $1.5$ M$_\odot$,  with  $\sim 56\%$ having a stellar companion and $\sim 40\%$ a planetary or brown dwarf companion. 
Roughly $10\%$ of all stellar companions are thought to be white dwarf companions.
While these fractions are reduced by roughly $10-20 \%$ during the AGB phase due to tidal inspiral and stellar mass loss, most stars with $\dot{M} \geq 10^{-7}$ \msunyr, as considered here, will have at least one (sub)stellar companion. \cite{Decin2020}


Unlike a planetary companion, a stellar companion's UV radiation field initiates UV-driven chemistry in this otherwise shielded inner wind.
UV photons originating from either the AGB star itself\cite{VandeSande2019a} or a companion star\cite{VandeSande2022} are included in the reaction network as additional photoreactions.
UV fluxes are approximated by blackbody radiation, set by the temperature of the AGB star, T$_*$, or the temperature of the companion, \Tcomp, together with the radius of the companion, \Rcomp.
We consider three types of companion: a red dwarf (\Tcomp\ = $4000$ K, \Rcomp\ =$1.53\times 10^{10}$ cm), a solar-like star (\Tcomp\ = $6000$ K, \Rcomp\ = $8.14\times 10^{10}$ cm), and a white dwarf (\Tcomp\ = $10,000$ K and \Rcomp\ = $6.96 \times 10^{8}$ cm).
Unshielded photodissociation and photoionisation rates were calculating using cross sections where available, which were mainly taken from the Leiden Observatory Database\footnote{\url{https://home.strw.leidenuniv.nl/~ewine/photo/}}\cite{Heays2017}.
If not available, the rates were estimated by scaling the unshielded interstellar rate by the ratio of the integrated fluxes of the stellar or companion photons in the $912-2150$ \AA\ range.
The scaling factors and the corrections for overestimated photoionisation rates can be found in Van de Sande \& Millar\,(2022)\cite{VandeSande2022}.

Internal UV photons are diluted geometrically as well as extinguished by dust.
The visual extinction caused by dust experienced by internal photons, determined by the outflow density, porosity, and \Rdust, and the intensity of the radiation, determined by \Tcomp\ and \Rcomp.
The companion star is assumed to lie within the dust free region before \Rdust.

While the impact of AGB UV photons is limited because of their low blackbody temperatures\cite{VandeSande2019a}, that of companion UV photons can be substantial throughout the entire outflow.
Solar-like companions have the largest influence, whilst white dwarf companions have a smaller effect. 
The impact of red dwarf companions is not significant.
The result of the photochemistry induced by companion UV photons depends mostly on the extinction experienced by the photons and the intensity of the radiation.
In low extinction outflows, characterised by a low outflow density and/or a high porosity, photoreactions occur faster than two-body reactions.
This inhibits the reformation of parent species and chemistry among photoproducts, reducing the outflow to a mostly atomic and ionised state and making it appear apparently molecule-poor. 
In high extinction outflows, characterised by high outflow density and/or low porosity, two-body reactions occur faster than photoreactions, increasing the chemical complexity of the outflow. 
Daughter species that are formed in the outer regions can now be produced in the inner regions, resulting in a parent-like abundance profile which consists of a large inner region abundance followed by a gaussian decline due to photodissociation by interstellar UV photons in the outer envelope.

\begin{figure*}[ht]
 \centering
 \includegraphics[width=18cm]{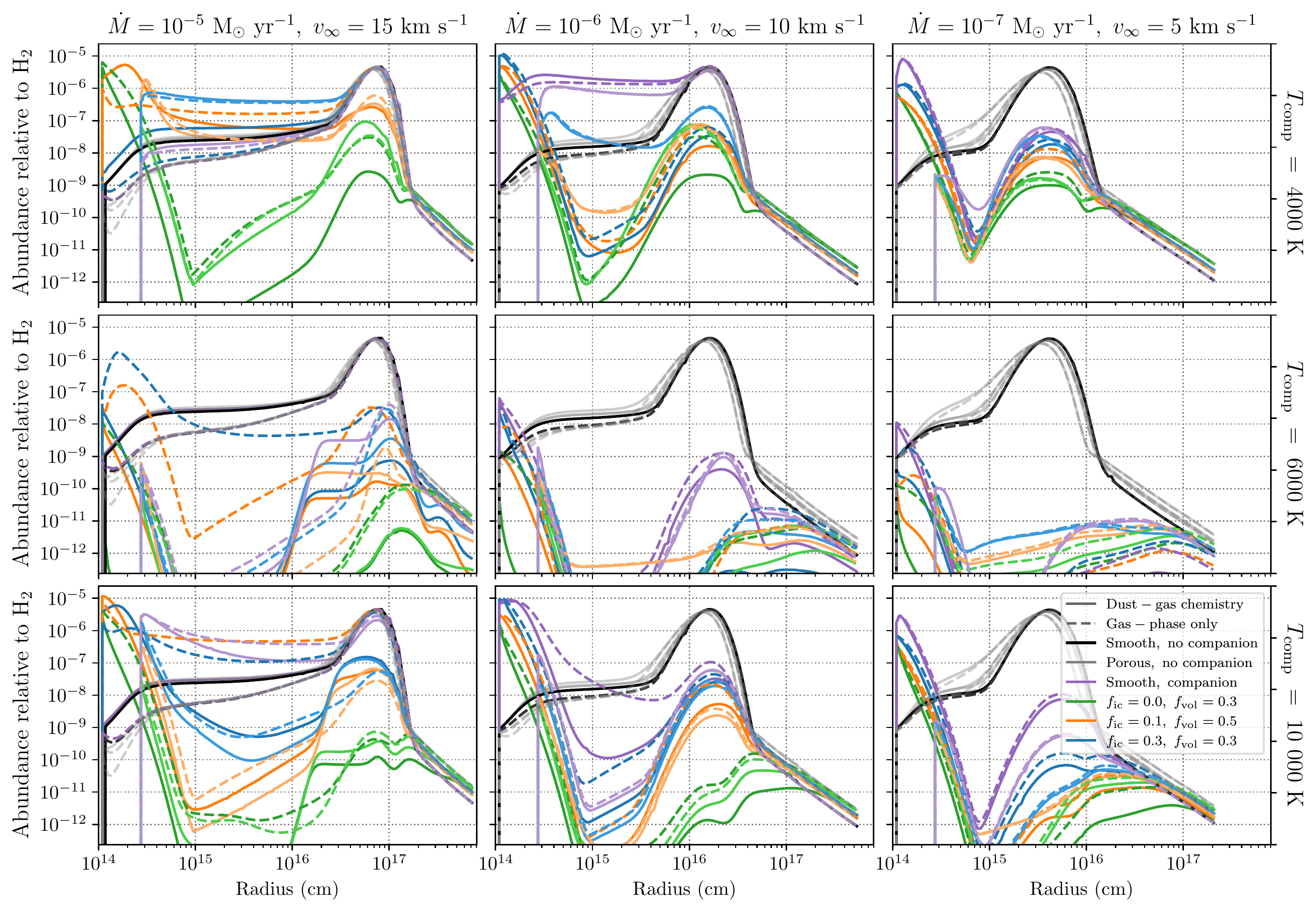}
 \caption{Fractional abundance of \ce{HS} relative to \ce{H2} for a selection of O-rich outflows with different outflow densities (columns) and companions (rows). 
 Different line styles show different chemistries included. Dashed lines: gas-phase chemistry only, solid lines: including dust-gas chemistry.
 Different colours show different density structures and the presence of a companion.
 Black: smooth outflow without a companion, gray: porous outflow without a companion.
 When including a companion, different shades correspond to the value of \Rdust. 
 Darker shade: \Rdust $= 2\ R_*$, lighter shade: \Rdust $= 5\ R_*$.
 Purple: smooth outflow with a companion. 
 Green: porous outflow with a companion, \fic $= 0.0$, \fvol $= 0.3$, $l_* = 10^{13}$ cm.
 Orange: porous outflow with a companion, \fic $= 0.1$, \fvol $= 0.5$, $l_* = 10^{13}$ cm.
 Blue: porous outflow with a companion, \fic $= 0.3$, \fvol $= 0.3$, $l_* = 5\times10^{12}$ cm.
}
 \label{fig:Orich-HS}
\end{figure*}

\subsection{Model limitations}			\label{subsect:model:limits}	

Several assumptions are made when introducing the physical and chemical complexities, limiting the one-dimensional model. 
As the porosity formalism assumes a stochastic mixture of clumps within an interclump medium, the clumps' specific configuration is not needed as input to the model.
While this is an advantage for model setup and exploration of the [\fvol, \fic, $l_*$] parameter space, reproducing specific (observed) outflows is not possible. 
Specific spherical asymmetries can only be approximated.

Dust nucleation is not included in the model: all dust is assumed to present at the start of the model with a fixed GSD. 
Hence, we assume that dust formation has taken place in the inner region before the start of the model and that GSD is kept constant throughout the calculations, as ices and refractory organic material are physisorbed rather than chemisorbed.

Since the same method of calculating photoreaction rates is used for companion and AGB UV photons, this implies that the companion is located at the centre of the star.
However, misplacing the companion by up to 5 R$_*$, the largest value of \Rdust\ (Table \ref{table:model}), is negligible compared to the scale of the outflow.
Additionally, the companion's radiation field is continuously present in our one-dimensional model. 
In Van de Sande \& Millar (2022)\cite{VandeSande2022}, we argue that occultation of the close-by companion by the AGB star likely has a limited effect on the chemistry, making our model a reasonable first-order approximation of the effects of a companion within the dust forming region.

\begin{figure*}[ht]
 \centering
 \includegraphics[width=18cm]{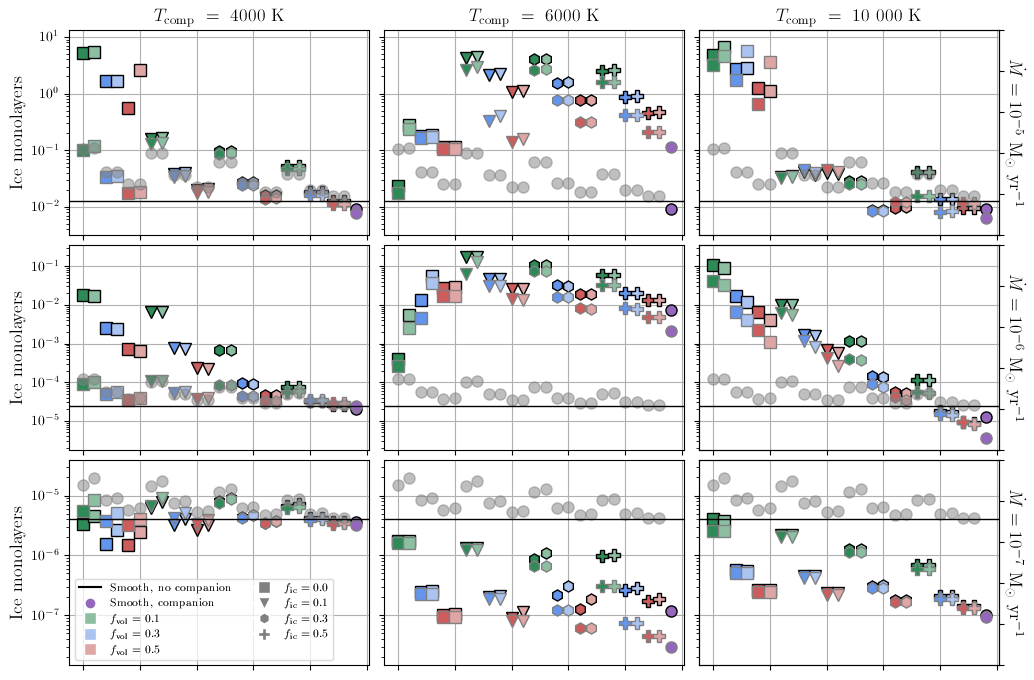}
 \caption{Number of ice monolayers at the end of the outflow for all calculated O-rich outflows with different density structures and companions. 
 The black solid line shows the number of monolayers in a smooth outflow without a companion.
 Different shapes and colours correspond to different density structures and the presence of a companion.
 Purple circles: smooth outflow with companion.
 Grey circles: porous outflows without a companion.
 The other shapes show the results for porous outflows with a companion. 
 The shape signifies the value of \fic: squares, triangles, hexagons, crosses correspond to \fic\ $= 0.0,\ 0.1,\ 0.3,\ 0.5$, respectively.
 The colour signifies the value of \fvol: green, blue, red correspond to \fvol\ $= 0.1,\ 0.3,\ 0.5$, respectively.
 Different shades signify different values of $l_*$. Lighter colour: $l_* = 5\times10^{12}$ cm, darker colour:  $l_* = 10^{13}$ cm.
 The edge colour of the shapes shows different values of \Rdust. Black edge: \Rdust\ = $2\ R_*$, grey edge: \Rdust\ = $5\ R_*$.
   }
 \label{fig:Orich-icecov}
\end{figure*}

\begin{figure*}[ht]
 \centering
 \includegraphics[width=18cm]{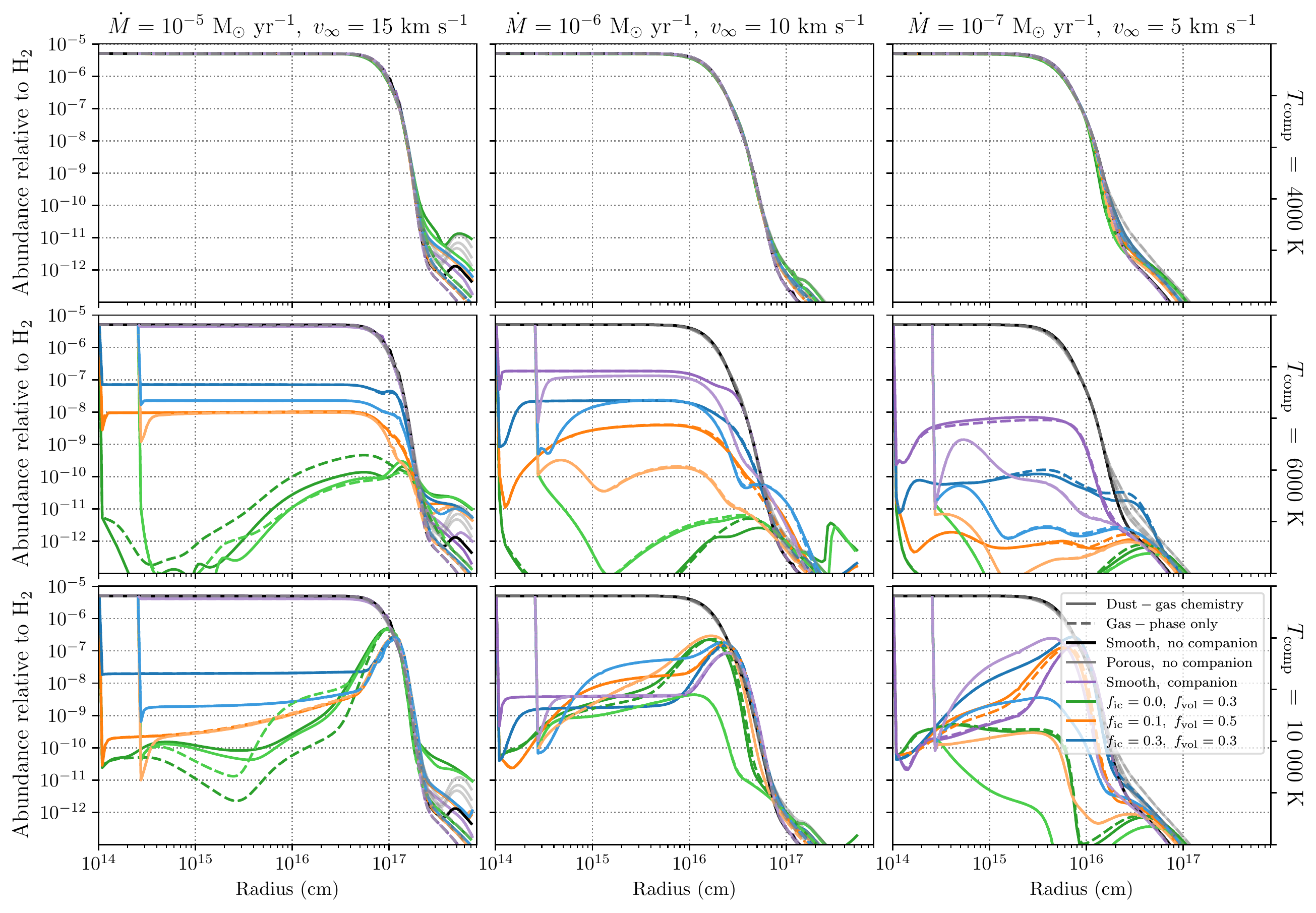}
 \caption{Fractional abundance of the parent \ce{SiO} relative to \ce{H2} for a selection of C-rich outflows with different outflow densities (columns) and companions (rows). 
 Different line styles show different chemistries included. Dashed lines: gas-phase chemistry only, solid lines: including dust-gas chemistry.
 Different colours show different density structures and the presence of a companion.
 Black: smooth outflow without a companion, gray: porous outflow without a companion.
 Including a companion, different shades correspond to the value of \Rdust. 
 Darker shade: \Rdust $= 2\ R_*$, lighter shade: \Rdust $= 5\ R_*$.
 Purple: smooth outflow with a companion. 
 Green: porous outflow with a companion, \fic $= 0.0$, \fvol $= 0.3$, $l_* = 10^{13}$ cm.
 Orange: porous outflow with a companion, \fic $= 0.1$, \fvol $= 0.5$, $l_* = 10^{13}$ cm.
 Blue: porous outflow with a companion, \fic $= 0.3$, \fvol $= 0.3$, $l_* = 5\times10^{12}$ cm.
}
 \label{fig:Crich-SiO}
\end{figure*}

%

\begin{figure*}[ht]
 \centering
 \includegraphics[width=18cm]{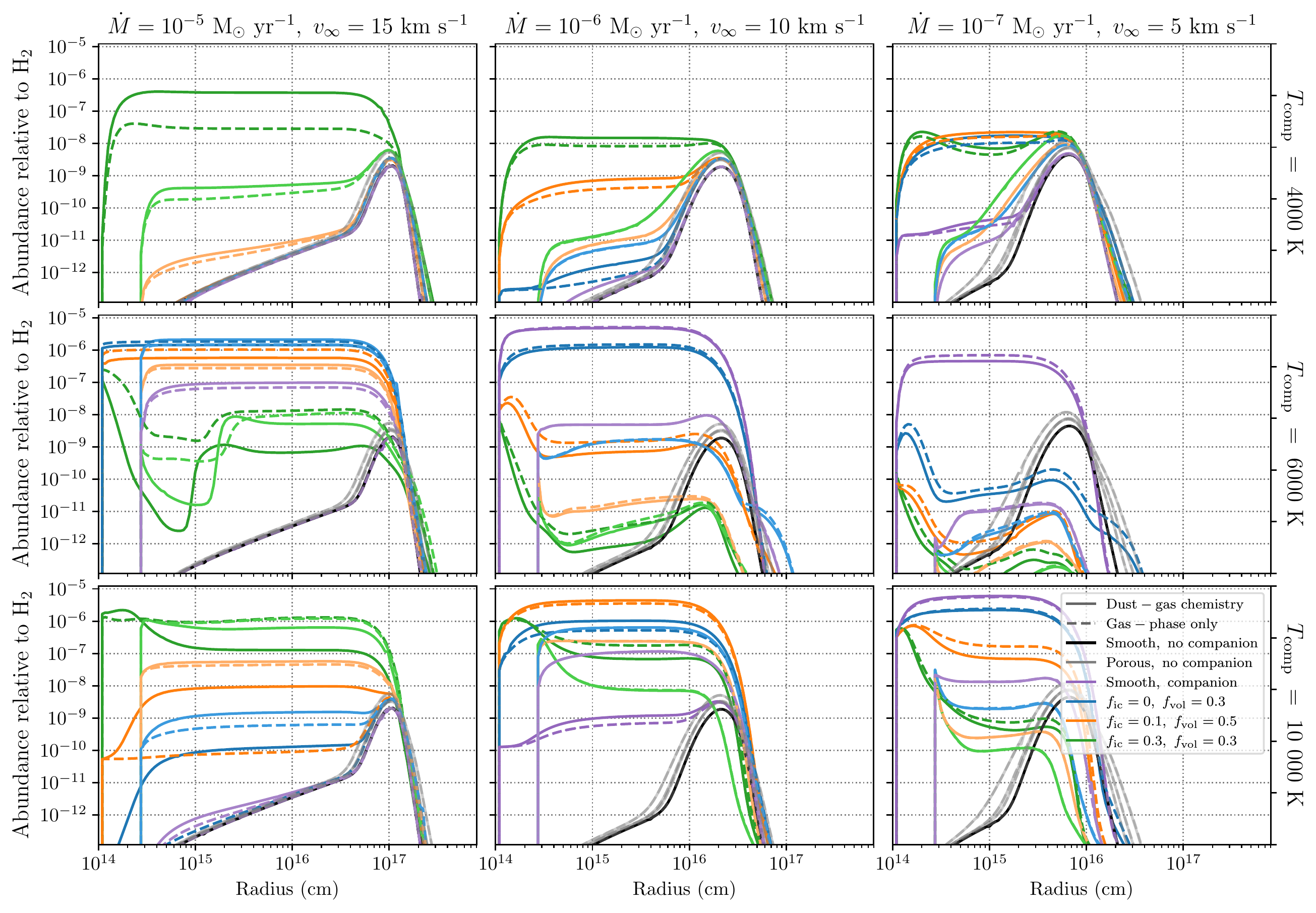}
 \caption{Fractional abundance of the \ce{H2CS} relative to \ce{H2} for a selection of C-rich outflows with different outflow densities (columns) and companions (rows). 
 Different line styles show different chemistries included. Dashed lines: gas-phase chemistry only, solid lines: including dust-gas chemistry.
 Different colours show different density structures and the inclusion of a companion.
 Black: smooth outflow without a companion, gray: porous outflow without a companion.
 When including a companion, different shades correspond to the value of \Rdust. 
 Darker shade: \Rdust $= 2\ R_*$, lighter shade: \Rdust $= 5\ R_*$.
 Purple: smooth outflow with a companion. 
 Green: porous outflow with a companion, \fic $= 0.0$, \fvol $= 0.3$, $l_* = 10^{13}$ cm.
 Orange: porous outflow with a companion, \fic $= 0.1$, \fvol $= 0.5$, $l_* = 10^{13}$ cm.
 Blue: porous outflow with a companion, \fic $= 0.3$, \fvol $= 0.3$, $l_* = 5\times10^{12}$ cm.
}
 \label{fig:Crich-H2CS}
\end{figure*}

\begin{figure*}[ht]
 \centering
 \includegraphics[width=18cm]{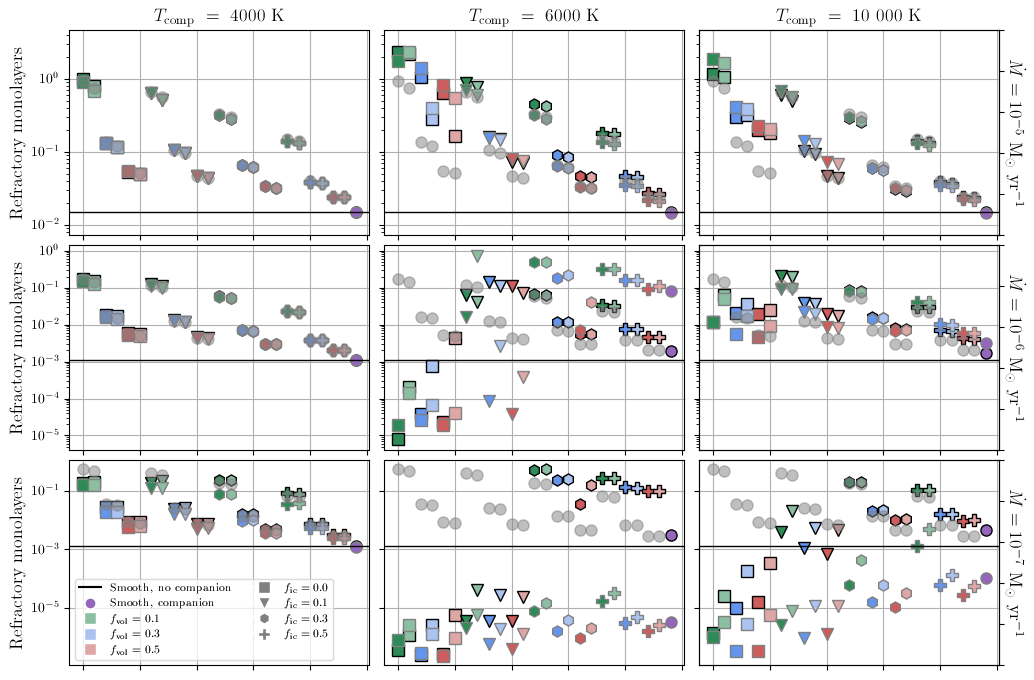}
 \caption{Number of refractory monolayers at the end of the outflow for all calculated C-rich outflows with different density structures and companions. 
 The black solid line shows the number of monolayers in a smooth outflow without a companion.
 Different shapes and colours correspond to different density structures and the presence of a companion.
 Purple circles: smooth outflow with companion.
 Grey circles: porous outflows without a companion.
 The other shapes show the results for porous outflows with a companion. 
 The shape signifies the value of \fic: squares, triangles, hexagons, crosses correspond to \fic\ $= 0.0,\ 0.1,\ 0.3,\ 0.5$, respectively.
 The colour signifies the value of \fvol: green, blue, red correspond to \fvol\ $= 0.1,\ 0.3,\ 0.5$, respectively.
 Different shades signify different values of $l_*$. Lighter colour: $l_* = 5\times10^{12}$ cm, darker colour:  $l_* = 10^{13}$ cm.
 The edge colour of the shapes shows different values of \Rdust. Black edge: \Rdust\ = $2\ R_*$, grey edge: \Rdust\ = $5\ R_*$.
  }
 \label{fig:Crich-refrcov}
\end{figure*}

\section{Results}				\label{sect:results}

Including different complexities in the chemical model can result in large differences compared to the fiducial gas-phase chemistry only, spherically symmetric outflow model.
Both the gas-phase chemistry throughout the outflow and the ice and refractory coverage of its dust grains can be severely affected.
We find that including companion UV photons has the largest effect, especially in combination with a porous outflow.
The specific influence depends on the outflow's density and porosity, as well as the type of companion included.

The influence of different (combinations of) complexities on the gas-phase composition and the dust's ice and refractory coverage at the end of the outflow is discussed in Sects \ref{sect:results:orich} and \ref{sect:results:crich} for O-rich and C-rich outflows, respectively.
The results shown in this Section all assume the MRN distribution for the GSD.
Any differences caused by changing the GSD are discussed in Sect. \ref{sect:disc}.
All fractional abundances are expressed relative to \ce{H2}, which follows a $r^{-2}$ distribution.

The ESI shows the fractional abundance profiles of all parent species and their daughters, along with all species that show a significant change in abundance when including a companion.
It also includes the number of monolayers of dust and ice for all assumed GSDs and the ice and refractory coverage at the end of the outflow for the GSDs different to the MRN distribution.


\subsection{O-rich outflows}			\label{sect:results:orich}

To illustrate the combined effects of the different complexities, we consider two different species: the parent SO, which can show a bumped abundance profile, and the daughter HS, whose abundance profile can change dramatically.


Figure \ref{fig:Orich-SO} shows the abundance of the parent SO in outflows with different densities, porosities, and companions.
Including porosity and dust-gas chemistry, separately or combined, does not affect its abundance profile significantly.
Including a companion only decreases the inner wind abundance in lower density outflows, creating a bumped abundance profile.
The largest effects are seen when combining a companion and porosity. 
For a red dwarf companion, the SO profile becomes bumped, with a decreased inner wind abundance by up to four orders of magnitude followed by a peak in abundance, as outflow density decreases and porosity increases.
Higher density outflows with a smooth or low porosity density distribution are not affected.
Including a white dwarf leads to similar profiles, with lower porosities now also resulting in a bumped profile in the higher density outflows, with a decrease in the inner wind abundance of up to some two orders of magnitude.
For $\dot{M} = 10^{-7}$ \msunyr, the peak of the bump decreases with increasing porosity.
In higher density outflows, a highly porous outflow gives rise to a wider bump, ranging from the inner wind to where SO is photodissociated by interstellar UV photons.

The abundance of the daughter HS is shown in Figure \ref{fig:Orich-HS}.
In the fiducial model, HS is present in a shell in the outer wind, where it is formed via photodissociation of the parent \ce{H2S}.
Including porosity alone leads to a widening of the shell, especially in lower density outflows.
Dust-gas chemistry alone has some influence on the inner wind abundance, increasing it by a factor of a few in higher density outflows.
Again, including a companion shows the largest influence.
Including a red dwarf or white dwarf changes the abundance profile to a high inner wind abundance followed by a sharp decline of at least five orders of magnitude and a bump in abundance in the outer wind. 
The bump's abundance decreases with increasing porosity, flattening the abundance profile.
In high density outflows with a low porosity, we find that the inner wind abundance of HS increases by up to two orders of magnitude, even making it appear parent-like in a smooth outflow with $\dot{M} = 10^{-6}$ \msunyr.
For a solar-like companion, HS is not efficiently reformed, lowering its abundance in the outer wind. 
In lower density, high porosity outflows, the abundance profile is flattened.
Dust-gas chemistry leads to a large difference of more than five orders of magnitude compared to gas-phase only chemistry in porous, high density outflows, with HS severely decreased in the inner regions when including dust-gas chemistry.

The number of ice monolayers covering the dust at the end of the outflow is shown in Figure \ref{fig:Orich-icecov}.
We take the end of the outflow where the \ce{H2} number density is equal to $10$ cm$^{-3}$.
In outflows with $\dot{M} = 10^{-5}$ \msunyr, porosity increases the ice coverage by up to an order of magnitude, from a coverage of $\sim 1\%$ to $\sim 10\%$. 
In lower density outflows, this ice coverage increases by up to an order of magnitude, but remains low: we find an ice coverage of up to $\sim 0.01\%$ for $\dot{M} = 10^{-6}$ \msunyr and $\sim 0.001\%$ for $\dot{M} = 10^{-7}$ \msunyr.
Including porosity increases the coverage of the dust by up to a monolayer, with the largest increases seen for smaller values of \fic\ as a smaller fraction of the envelope mass is in the interclump component for these outflows.
When adding a companion, an increase of up to five orders of magnitude and a decrease of up to two orders of magnitude is possible. 
The final coverage can increase to several monolayers for \mdot\ $= 10^{-5}$ \msunyr, and stays below $10\%$ and $0.001\%$ for \mdot\ $= 10^{-6}$ and $10^{-7}$ \msunyr.

\subsection{C-rich outflows}			\label{sect:results:crich}

We consider again two species to showcase the combined effects of the different complexities: the parent SiO, whose abundance decreases when including a solar-like companion and can show a bumped profile for a white dwarf, and the daughter \ce{H2CS}, which can become observable for specific combinations of complexities.


Figure \ref{fig:Crich-SiO} shows the abundance of the parent SiO in outflows with different densities, porosities, and companions.
Porosity and dust-gas chemistry, combined or separately, do not have a significant impact on the chemistry, neither does including a red dwarf companion.
Including a solar-like companion leads to a decreases in abundance as porosity increases, with smooth outflows showing a decreases of 2 to 3 orders of magnitude in lower density outflows.
When including a white dwarf companion, the profile changes to a bumped profile, similar to SO in O-rich outflows.
The inner wind abundances decreases by 2 to 4 orders of magnitude, depending on the outflow's density and porosity, with a bump in abundance around $10^{-7}$ in the outer wind.
The abundance of the bump decreases with increasing porosity in lower density outflows.

%

Figure \ref{fig:Crich-H2CS} shows the abundance of \ce{H2CS}.
We find that including porosity increases the peak in abundance up to a factor of five and widens the peak in the lower density outflows.
Including a red dwarf leads to parent-like abundance profiles with increasing porosity.
For a solar-like star, this occurs with decreasing porosity. 
In lower density outflows, the abundance profile flattens with increasing porosity.
Including a white dwarf leads to a mix of the two previous effects: at high outflow densities, \ce{H2CS} becomes parent-like with increasing porosity, at low densities with decreasing porosity.
In the outflow with $\dot{M} = 10^{-7}$ \msunyr, a high porosity results in a large inner wind abundance followed by a steep decline.

The number of refractory monolayers covering the dust at the end of the outflow is shown in Figure \ref{fig:Crich-refrcov}.
The dust's refractory coverage in C-rich outflows with $\dot{M} = 10^{-5}$ \msunyr\ increases with porosity from $\sim 1.5\%$ in the smooth outflow up to two orders of magnitude to a complete grain coverage of $1$. 
In lower density outflows, the refractory coverage is increased by up to three orders of magnitude with porosity, reaching coverages of $\sim 10\%$.
In C-rich outflows, a companion can lead to an increase in refractory coverage of up to one order of magnitude and a decrease of up to two orders of magnitude at low outflow densities.
Nevertheless, the coverage stays below $10\%$, $1\%$, and $0.01\%$ in outflows with \mdot\ $= 10^{-5}$, $10^{-6}$, and $10^{-7}$ \msunyr, respectively.

\section{Discussion}			\label{sect:disc}

The chemical composition throughout the outflow can be strongly influenced when including different layers of complexity. 
While including a companion in a porous outflow has the largest effects, the other (combinations of) complexities show different signatures in the gas-phase and the ice and refractory coverage of the dust.
In Sect. \ref{subsect:disc:interplay}, we discuss the effects of the different complexities separately as well as the interplay between them. 
In Sect. \ref{subsect:disc:tracers}, we identify sets of gas-phase molecules that trace the dust-gas chemistry as well as the presence of a companion in a (porous) outflow.


\subsection{Interplay of complexities}			\label{subsect:disc:interplay}

Separately, porosity and dust-gas chemistry have a small effect on the overall chemistry of the outflow.
When including only \emph{porosity}, e.g. caused by a planetary companion that induces a large-scale structure in the outflow, the envelopes of parent species become mainly smaller with increasing outflow porosity, but the reduction in size is too small to be observed. 
The abundance profile of certain parents changes with increased porosity and lower outflow density: \ce{SO} in O-rich outflows shows a small bump before the onset of photodissociation, and the envelope of \ce{CS} in O-rich outflows and \ce{C2H4} in C-rich outflows become smaller with the knee in the outer wind widening.
As porosity increases, peak abundances of daughter species increase and shift closer to star, but this is effect is again likely too small to be observed.

Including \emph{dust-gas chemistry} does not lead to significant depletion of parent species for dust grains following the canonical MRN GSD and the GSD with larger grains than MRN.
Assuming silicates without Fe only likely leads to more significant depletion in O-rich outflows \cite{VandeSande2019b}.
However, most O-rich outflows host a mixture of dust grains, including warmer grains that have Fe inclusions \cite{Heras2005}, leading to smaller depletion levels.
Parent species are depleted onto dust assuming a GSD with smaller dust grains because of the larger number density of grain surface sites per unit volume.
For dust grains following the GSD with smaller dust grains, parent species with high binding energies (e.g., SiO, SiS, \ce{H2O}) are depleted onto dust grains in outflows with $\dot{M} = 10^{-5}$ \msunyr.
In O-rich outflows, the depletion takes place around $2\times 10^{16}$ cm, in C-rich outflows around $4\times 10^{16}$ cm.
Certain species, such as \ce{SiH4}, are efficiently formed via surface chemistry. 
These gas-phase tracers of dust-gas chemistry are released into the gas phase in the outer wind, forming a molecular shell (Sect. \ref{subsubsect:disc:tracers:dust}).
Assuming a GSD with smaller grains increases the abundance of the gas-phase tracers by about an order of magnitude.

Including a \emph{stellar companion} alone has the largest effect on the chemistry.
For smooth outflows, its effect is limited to lower density outflows (corresponding to lower extinction outflows) with a solar-like or white dwarf companion.
A red dwarf companion does not significantly impact the chemistry. 
Sub-stellar brown dwarfs are therefore not expected to have a significant impact.

When combining complexities, the impact on the dust- and gas-phase chemistry becomes more apparent. 
Porosity together with a stellar companion in a gas-phase only chemistry chemical model has been described in \citet{VandeSande2022} (see Sect. \ref{model:chem:companion}). 

\subsubsection{Porosity and dust-gas chemistry}

Combining porosity and dust-gas chemistry, we find that a larger porosity increases the abundance of the gas-phase tracers and the grain surface coverage.
The ice coverage in higher density O-rich outflows, where \ce{H2O} is the main ice component, is expected to be observable with JWST considering the ISO detections \cite{Omont1990,Justtanont1992}.
The abundances of most gas-phase tracers can increase up to an order of magnitude (Sect. \ref{subsubsect:disc:tracers:dust}).
Assuming a GSD with smaller grains than MRN leads to an additional increase by an order of magnitude. 


The coverages are not significantly changed when assuming a GSD with larger dust grains than MRN.
When assuming smaller dust grains, the ice and refractory coverage in both outflows decreases by a factor of a few, with porosity leading to a smaller possible increase.
While gas-phase species are now depleted, leading to a larger total abundance of ice and refractory material, the larger number density of dust surface sites per unit volume leads to a smaller number of monolayers.

\subsubsection{Porosity, dust-gas chemistry, and a stellar companion}

When combining all three complexities, we find that including a stellar companion has the largest effects on the overall chemistry of the outflow, significantly impacting both the gas phase composition and grain surface coverage.
The effects on the gas phase are similar to our previous results, which included gas-phase chemistry only \cite{VandeSande2022}: outflows where companion UV photons experience a low extinction (low outflow density and/or high porosity) appear molecule poor, while outflows with a high extinction (high outflow density and/or low porosity) show an increase in chemical complexity, with daughter species showing a parent-like abundance profile.

\paragraph*{Influence of dust-gas chemistry on the gas phase}

Including dust-gas chemistry generally does not significantly affect the inner wind abundances in outflows with an increased chemical complexity. 
These newly formed daughters are not depleted onto dust.
While some accretion does occur, most of the accreted ices are rapidly thermally desorbed or photoprocessed into refractory organic material if the ice is already complex.
Any changes to the gas-phase are predominantly caused by the availability of feedstock material.
As parents are depleted onto dust grains, this lowers the abundance of their photodissociation products in the gas phase and hinders the formation of more complex daughters in the inner wind. 
A notable example is HS in both C-rich and O-rich outflows which shows several orders of magnitude difference, rendering it effectively unobservable when including dust-gas chemistry.
This is linked to depletion of the parent CS: photodissociation by internal photons of CS is an important source of S, removing it from the gas phase reduces the available S for HS formation, and for \ce{H2S} formation as well in C-rich outflows.

Including a companion generally increases the abundance of the gas-phase tracers of dust-gas chemistry, present in a shell in the outer wind.
We find increases of about two orders of magnitude compared to a smooth outflow and of about one order of magnitude compared to clumpy outflows.
However, certain species are destroyed by including a companion (see Sect. \ref{subsect:disc:tracers}).
Assuming a GSD with smaller dust grains than MRN leads to a further increase by about an order of magnitude.

\begin{table*}[ht]
	\caption{Molecular tracers of different companions and porosities in C-rich outflows. If not specified, the tracer is valid for all densities and porosities.
	} 
    \centering
    \begin{tabular}{l r l l l }
    \hline  
    \noalign{\smallskip}
        ~ & ~ & Red dwarf & Solar-like & White dwarf  \\  
    \noalign{\smallskip}
    \hline
    \noalign{\smallskip}
    \multicolumn{2}{c}{Parents} & & & \\
    \noalign{\smallskip}
    \hline
    \noalign{\smallskip}
        (i) & $\searrow$  in abundance with $\nearrow$ porosity  & \ce{C2H4}, \ce{NH3} & All except HCN & All except HCN   \\ 
        ~ & ~ & ~ & ~ & High porosity: \ce{C2H2}, \ce{CH4},    \\ 
        ~ & ~ & ~ & ~ & \quad CS, \ce{NH3}, SiS    \\ 
        ~ & $\nearrow$ in abundance with $\nearrow$ porosity  & Low porosity: \ce{H2S} & ~ & \ce{H2O}  \\ 
    \noalign{\smallskip}
        (ii) & bump for $\nearrow$ porosity & High porosity: \ce{H2S} & ~ & Low density: SiO  \\ 
        ~ & bump for $\searrow$ porosity & ~ & High density: \ce{H2S} & High density: SiO  \\ 
    \noalign{\smallskip}
    \hline
    \noalign{\smallskip}
    \multicolumn{2}{c}{Daughters} & & & \\
    \noalign{\smallskip}
    \hline
    \noalign{\smallskip}
        (iii) & $\nearrow$ inner wind abundance with $\nearrow$ porosity & \ce{CH3}, HS, NH,  & Low density: HS & High density: \ce{CH3}   \\ 
        ~ & ~ & \ce{NH2}, \ce{C6H}  & \ce{C2H} & High density: \ce{C6H}  \\ 
        ~ & ~ & \ce{H2CS}, HCSi, NS & High density: HCSi &   \\ 
    \noalign{\smallskip}
        ~ & with $\searrow$  porosity & ~ & High density: \ce{CH3},  & Low density: \ce{CH3}   \\ 
        ~ & ~ & ~ &  \quad CN, OH, HS &    \\ 
        ~ & ~ & ~ & ~ & High density: NH, HS   \\ 
        ~ & ~ & ~ & \ce{HC3N}, \ce{HC5N}, \ce{HC7N} & Low density: \ce{C6H}  \\ 
    \noalign{\smallskip}
        (iv) & parent-like with $\nearrow$ porosity & ~ & High density: SiN, SiC, SiNC & SiNC  \\ 
    \noalign{\smallskip}
        ~ & with $\searrow$  porosity & HS & \ce{CH3CN}, \ce{H2CS}, NS,  & Low density: \ce{CH3CN},   \\ 
        ~ & ~ & ~ &  Low density: SiN, SiC, SiNC & \quad \ce{H2CS}, NS, SiC, SiN \\ 
    \hline 
    \end{tabular}%
    \label{table:tracers-crich}    
\end{table*}

\paragraph*{Influence of a companion on the ice coverage}

The influence of a companion on the ice coverage of the dust depends mainly on surface chemistry.
When including a companion, the abundance of atoms in the gas phase and on the dust increases in the outer region.
The atoms are highly reactive on the dust surface, efficiently (re)forming, e.g., ice \ce{H2O}, \ce{H2S}, HCN, \ce{NH3}, and SO, and increasing the ice coverage.
In both O-rich and C-rich outflows, the final ice coverage mainly depends on \ce{H2O} formation on the surface, linked to the availability of H. 
This leads to a larger coverage compared to the fiducial dust-gas chemistry model for \mdot\ $= 10^{-6}$ \msunyr with a solar-like or white dwarf companion.

In C-rich outflows, the reactivity of the surface chemistry is the main cause for an increase in final ice coverage.
The increase mainly takes place in low extinction outflows, where surface chemistry efficiently forms \ce{H2O} ice.
Surface chemistry also has an indirect effect on the ice coverage. 
In outflows with \mdot\ $= 10^{-5}$ \msunyr, the peak abundance of SiS ice can increase by an order of magnitude with the peak broadening by half a magnitude in both O-rich and C-rich outflows. 
SiS ice is only formed via accretion from the gas phase.
The gas-phase abundance of SiS increases in this region via the reaction \ce{ SiH2+ + S -> HSiS+ }, which dissociatively recombines with electrons to form SiS.
\ce{SiH2+} is produced by photoionisation of \ce{SiH2}, formed on the surface and released into the gas phase by photodesorption.

%
%

\paragraph*{Influence of a companion on the refractory coverage}

In general, a companion does not lead to a larger range of possible coverages compared to including porosity alone.
However, it can affect the refractory coverage of individual porous outflows.
In high extinction  C-rich outflows, the complexity of the gas phase and ices is reduced, leading to a smaller refractory coverage.
In low extinction outflows, a solar-like companion can lead to an increase in coverage of up to two orders of magnitude, a white dwarf to up to one order of magnitude, and a red dwarf to a factor of a few.
The observability of the refractory material with JWST depends on its composition.
As refractory material is assumed to be chemically inert in our model, estimates are difficult to make.
For O-rich outflows, a companion does not significantly increase the coverage of any specific outflow.
A solar-like companion significantly reduces the coverage because of its large effect on the gas phase, rendering the dust effectively bare.
Nonetheless, the refractory coverage of O-rich dust is low, with a negligible coverage for all outflow densities.


\subsection{Chemical tracers of outflow structure}			\label{subsect:disc:tracers}

Specific (combinations of) species allow us to constrain the underlying physics and chemistry at play in the outflow. 
The effects of porosity only on the chemistry is too small to constrain.
While dust-gas chemistry can lead to a significant ice and/or refractory coverage of the dust grains, these are not easily observable, although observations of AGB envelopes with JWST may reveal clues on the dust composition and presence or otherwise of ices.
We identify gas-phase tracers of active dust-gas chemistry in O-rich and C-rich outflows in Sect. \ref{subsubsect:disc:tracers:dust}.
The presence of a companion, along with whether the outflow is highly porous or not, can be constrained as well, allowing us to distinguish between planetary and stellar companions shaping the outflow.
Sect. \ref{subsubsect:disc:tracers:dust} discusses the combinations of species for O-rich and C-rich outflows.

\begin{table*}[ht]
	\caption{Molecular tracers of different companions and porosities in O-rich outflows. If not specified, the tracer is valid for all densities and porosities.
	} 
    \centering
    \begin{tabular}{l r l l l }
    \hline  
    \noalign{\smallskip}
        ~ & ~ & Red dwarf & Solar-like & White dwarf  \\ 
    \noalign{\smallskip}
    \hline
    \noalign{\smallskip}
    \multicolumn{2}{c}{Parents} & & & \\
    \noalign{\smallskip}
    \hline
    \noalign{\smallskip}
        (i)  & $\searrow$ in abundance with $\nearrow$ porosity  & SiS, \ce{CO2}, CS,  & CS, HCN, \ce{NH3}, & CS, HCN, \ce{NH3},  \\ 
        ~ & ~ & \ce{NH3}, \ce{H2S} & SiO, SiS, \ce{H2O}, & SiS, \ce{H2S}  \\ 
        ~ & ~ & ~ & CO2 (low porosity) &   \\ 
    \noalign{\smallskip}
         & $\nearrow$ in abundance with $\nearrow$ porosity  & ~ & \ce{CO2} (high porosity) &   \\ 
    \noalign{\smallskip}
        (ii) &  bump for $\nearrow$ porosity & SO, \ce{SO2} & SO, \ce{SO2} & SO, \ce{SO2}, SiO  \\ 
    \noalign{\smallskip}
    \hline
    \noalign{\smallskip}
    \multicolumn{2}{c}{Daughters} & & & \\
    \noalign{\smallskip}
    \hline
    \noalign{\smallskip}
       (iii) & $\nearrow$ inner wind abundance with $\nearrow$ porosity & Low density: HS & CN & CN, OH,  \\ 
       ~ &  ~ & NO, \ce{N2O} & ~ & NO, \ce{N2O}  \\ 
        ~ & with $\searrow$ porosity & ~ & ~ &   \\ 
    \noalign{\smallskip}
        ~ & parent-like with $\nearrow$ porosity & High density: HS, NS & Low density: OH & High density: SiN, NS  \\ 
        ~ & with $\searrow$ porosity & ~ & NS, SiN, \ce{N2O} &   \\ 
    \noalign{\smallskip}
        (iv) &Flattening with $\nearrow$ porosity & ~ & ~ & HS, SiC, SiN, NS \\ 
    \hline 
    \end{tabular}%
    \label{table:tracers-orich}    
\end{table*}

\subsubsection{Dust-gas chemistry tracers}				\label{subsubsect:disc:tracers:dust}

When including dust-gas chemistry, \ce{SiH2}, \ce{SiH3}, \ce{SiH4}, and \ce{H2SiO} appear in a shell around $10^{17}$ cm in both O-rich and C-rich outflows with $\dot{M} = 10^{-5}$ \msunyr.
In O-rich outflows, \ce{SiH2}, \ce{SiH3}, and \ce{SiH4} have peak abundances of around $10^{-9}$ and \ce{H2SiO} has one of $\sim 5\times 10^{-10}$.
In C-rich outflows, this decreases to $10^{-11}$ for the silicon hydrides and $\sim 5\times 10^{-13}$ for \ce{H2SiO}.

Their peak abundances depend on the presence of other complexities.
When porosity is included, they increase by roughly an order of magnitude.
Including a companion leads to an increase of two orders of magnitude, which can increase by an additional order of magnitude when combining it with porosity.

O-rich outflows have two additional tracers: \ce{NH2OH} and \ce{H2O2}.
\ce{NH2OH} reaches abundances of $5\times 10^{-9}$ around $3\times 10^{17}$ cm. 
Its abundance decreases by an order of magnitude when including a solar-like companion, decreasing further when increasing the outflow's porosity.
\ce{H2O2} is present in a broad shell from about $10^{16}$ to $10^{17}$ cm, with a peak abundance of a few times $10^{-7}$.
The shell becomes more narrow when including a companion, shifting to $10^{17}$ cm. 
Highly porous outflows decrease its abundance by an order of magnitude; in the case of a solar-like companion, its abundance is drastically reduced to $10^{-11}$.

\subsubsection{Companion and porosity tracers}			\label{subsubsect:disc:tracers:comp}

The presence of a companion star, especially if the outflow is porous, can be traced by a combination of parent species, daughter species, and uncommon species formed via inner wind photochemistry.
The sets of species depend on the type of chemistry, the outflow's porosity, and the type of companion.

We find that the chemistry in the inner region is mainly driven by the competition between O and S for the newly liberated C in the inner wind.
A large number of the tracer species therefore contain sulphur. 
Newly liberated N in the inner wind also leads to large abundances NS, SiN, and NO.

\paragraph*{C-rich outflows}

Table \ref{table:tracers-crich} lists all molecular tracers for different types of companions, densities, and porosities in C-rich outflows.
We distinguish different types of behaviour: for parents (i) a decrease or increase in parent abundance with increasing porosity, (ii) the appearance of a bump in the abundance profile with increasing or decreasing porosity; for daughters, (iii) a larger inner wind abundance followed by a peak in abundance as porosity increases or decreases, and (iv) the abundance profile changing to a parent-like profile with a large inner wind abundance followed by a gaussian decline as porosity increases or decreases.

We find that the parents CO, \ce{N2}, and HCN are unaffected by the presence of a companion.
A solar-like companion leads only to smaller envelope size of HCN in high porosity, lower density outflows.
The parent \ce{H2S} is most strongly affected, its abundance profile changing to a steep decline followed by a bump, whose peak abundance decreases as the outflow density decreases, in all low density outflows and in high porosity, high density outflows.
For SiO, including a white dwarf changes its abundance profile to a lower inner wind abundance followed by a bump in higher density outflows and to a shell-like profile in lower density outflows, with the abundance peak decreasing with increasing porosity.

The daughter HS follows the shape of the \ce{H2S} profile.
The abundance profile of SO is sensitive to the type of companion and porosity of the outflow.
For a red dwarf companion, the peak in abundance of SO in the outer wind widens with increasing porosity, creating a broader shell. 
For a solar-like and white dwarf companion, SO becomes parent-like with increasing porosity in high density outflows, and with decreasing porosity in lower density outflows.

The behaviour of the cyanopolyynes and hydrocarbon radicals depends on outflow density and porosity for a solar-like and white dwarf companion.
In low density, high porosity outflows, they show dramatically smaller abundances.
For a solar-like companion, we find that the inner wind abundance of \ce{HC3N}, \ce{HC5N}, and \ce{HC7N} increases with decreasing porosity. 
As porosity increases, their profiles change to a broad shell.
The peak abundance drastically decreases in lower density outflows, rendering the longer chains unobservable in low density, high porosity outflows.
The inner wind abundance of \ce{C2H} increases in high density, high porosity outflows. 
For lower densities, the profile is flattened with increasing porosity, where the inner wind abundance increases while the peak abundance decreases.
The profiles of \ce{C4H} and \ce{C6H} change to a broad peak, again disappearing in low density, high porosity outflows.
For a white dwarf companion, the inner wind abundance of \ce{HC3N}, \ce{HC5N}, \ce{HC7N}, \ce{C2H}, and \ce{C4H} increases with increasing porosity, with their profiles flattening in lower density outflows. 
The inner wind abundance of \ce{C6H} increases with increasing porosity in higher density outflows.
In lower density outflows, the profile flattens with increasing porosity.

\paragraph*{O-rich outflows}

Table \ref{table:tracers-orich} lists all molecular tracers for different types of companions, densities, and porosities in O-rich outflows.
We distinguish the same types of behaviour as for the C-rich outflow.

We find that the parents CO and \ce{N2} are unaffected by the presence of a companion.
\ce{H2S} is efficiently destroyed when including a solar-like companion, rendering it apparently absent in lower density outflows and strongly reducing its abundance in high density outflows.
In high porosity outflows with a red dwarf or white dwarf companion, the abundance profile of \ce{H2S} changes to a high inner wind abundance followed by a bump.
In low porosity outflows, it has a smaller abundance, but parent-like shape.
The abundance profiles  of SO and \ce{SO2} changes to a lower initial abundance followed by a bump as porosity increases, becoming shell-like for lower densities.
The envelope size of \ce{H2O} and HCN becomes smaller in low density, high porosity outflows when including a white dwarf companion.

\section{Conclusions}			\label{sect:conclusions}

We presented the results of the most physically and chemically complex AGB chemical kinetics model to date, including porosity, dust-gas chemistry, and companion UV photons.
The chemical validity of the model is extended from the outer wind into the dust interaction zone, while being able to include the effects of spherically symmetric outflows, produced by a stellar or substellar companion.
The effects of each of these complexities was studied separately in previous work.
Here, we combined them all into one model and considered their interdependencies.

Porosity alone does not significantly impact the gas-phase composition of the outflow. 
It increases the ice and refractory coverage of the dust grains, with the specific increase depending on the chemistry of the outflow and the material considered.

Dust-gas chemistry does not significantly impact the gas-phase composition, except when a grain size distribution with large number of dust surface sites is assumed. 
However, certain gas-phase species, such as \ce{SiH4}, \ce{H2SiO}, and \ce{N2O}, are efficiently formed via surface chemistry and released into the gas phase via photodesorption by interstellar photons in the outer wind, forming a molecular shell. 
The abundance of this shell increases up to an order of magnitude when including porosity.
Including a companion increases the abundance of \ce{SiH4} and \ce{H2SiO}, but decreases that of \ce{N2O}.

We find that a companion star combined with a porous outflow, with or without dust-gas chemistry, shows the largest effect.
The impact on the gas-phase chemistry depends on the outflow density, porosity, and the assumed onset of dust formation, as well as the intensity of the companion's radiation.
Depending on the configuration, certain parent and daughter species are efficiently destroyed, with other daughters showing an increased inner wind abundance or even parent-like behaviour.
We have identified molecular tracers that allow us to discern the type of companion and porosity of the outflow for both O-rich and C-rich outflows.
Solar-like companions show the largest effects and can be inferred using our tracer molecules.  
The effect of white dwarfs is smaller, but their presence can be inferred as well.
Red dwarf companions can be inferred only in high porosity outflows. 
In this case, specific outflow structures are needed for chemistry to be used as a tool to distinguish between stellar and substellar companions.

Our results confirm that chemistry can be used to distinguish between a stellar or substellar companion shaping the outflow and constrain the underlying density distribution.
Even though the new model has more input parameters, it is possible to constrain them based on observations, as long as a suite of molecules is targeted. 
This makes the model an important tool to help interpret and guide observations.

While the chemical model is the most complex to date, it does not cover the chemistry of the entire outflow.
In order to achieve such a model, dust formation and growth need to be included. 
Additionally, to accurately model the chemistry in outflows with specific observed density distributions, such as spirals and disks, appropriate parameterisations or hydrodynamical modelling output are required.
Nevertheless, this model is a important first step in the development of such three-dimensional, full outflow chemical models.


%
%

\section*{Author Contributions}
Marie Van de Sande: conceptualisation, methodology, software, validation, investigation, visualisation, writing - original draft.
Catherine Walsh: conceptualisation, software, investigation, writing - review \& editing.
Tom J. Millar: conceptualisation, software, investigation, writing - review \& editing.

\section*{Conflicts of interest}
There are no conflicts to declare.

\section*{Acknowledgements}
MVdS acknowledges support from the  European Union's Horizon 2020 research and innovation programme under the Marie Skłodowska-Curie grant agreement No 882991. 
C.W. acknowledges financial support from the University of Leeds and from the Science and Technology Facilities Council (STFC, grant numbers ST/T000287/1 and MR/T040726/1)
TJM gratefully acknowledges the receipt of a Leverhulme Emeritus Fellowship and the STFC for support under grant numbers ST/P000312/1 and ST/T000198/1.



\balance


\bibliography{chemistry} 
\bibliographystyle{rsc} 

\end{document}